# Orbits and Individual Masses of Some Visual Binaries

Essam A. Elkholy[1,2], Waleed H. Elsanhoury[2,3] and Mohamed I. Nouh[2]

[1]Physics Department, College of Science, Northern Border University, Arar, Saudi Arabia

[2]Astronomy Department, National Research Institute of Astronomy and Geophysics (NRIAG), 11421 Helwan, Cairo, Egypt (Affiliation ID: 60030681)

[3]Physics Department, Faculty of Arts and Science, Northern Border University, Turaif Branch, Saudi Arabia

**Abstract:** The orbits of visual binary systems still attract many of the working groups in astronomy. These orbits are the most important and reliable source of the stellar masse. In the present paper, we are going to compute orbits and dynamical masses of some visual binaries using an independent code. We used Kowalsky method to compute the geometrical elements. The dynamical elements (the period and the time of the periastron passage) are computed by implementing the double areal constant. We used the developed code to calculate the orbits for the four visual binaries, WDS J02262+3428, WDS J14310-0548, WDS J17466-0354, and WDS J12422+2622. We introduced a new orbit of the neglected visual binary WDS J17466-0354 and modified orbits for the rest three binaries. Using the Gaia DR2 parallaxes we computed the total masses of the systems. Comparing the adopted total masses with those derived from the mass spectral type relation revealed good agreement.

1. Introduction

The study of visual binaries is one of the most crucial sources of the current knowledge of stellar masses is of great importance. Besides, the utilization of these masses leads to the discovery of the relation of mass-luminosity, which in turn supports many theories of stellar evolution. On the other hand, the correlation between binary parameters provides important data for star formation theory. The problem of computing orbital elements of a binary from an observed set of positions is the determination of visual binary orbits. Many authors suggested automated orbit determination techniques, given modern computers. In the same manner as the Thiele-Innes-van den Bos method [1-3] employed a $3\frac{1}{2}$ observing points method. All observations are used simultaneously only in the final stage.

A sufficiently accurate set of orbital parameters is required for the first approximation by many techniques or methods proposed. Eichhorn and Xu [4] have thus far developed a method that requires a precise initial orbit and only in the improvement phase are all the observations used at the same time. Catovic and Olevic [5] suggested a falsified observation approach to the solution of the least-squares, which should be chosen as an ellipse. As a result, you can create many elliptical orbits by changing the position, and the final decision to choose the best orbit is left to the computer. Pourbaix [6] and Pourbaix and Lampens [7] have developed a method based on a function to quantify distance from the position observed to the position calculated. The simulated annealing method has been used to minimize it successfully, thus minimizing the function of the best orbit. Nouh et al. [8] introduce an algorithm implementing an optimum point ($\rho_a, \theta_a$) that minimizes the average length of a particular function between the smallest and the least square one.

Our aim of the present paper has two folds, the first is to introduce computational algorithms for the determination of the visual orbits using Kowalsky method. The second is to implement the algorithm to determine the orbital elements and individual masse of some selected visual binary systems. To declare the efficiency of the calculations, we predict the ephemerides of the systems and compare the results with other already computed orbits.

The structure of the paper is as follows. Section 2 is devoted to the computational method used to calculate the orbits. In section 3 we draw the orbital elements of the visual systems under study. Section 4 concerned with the computational algorithm. Section 5 devoted to results. In section 6 we outlined the conclusion.

## 2. Computational Method

The word orbital elements are extracted from the observation of celestial motions. To characterize the movement of the components of the visual binaries in their orbit, seven quantities are required, the orbital period $P$ (usually expressed in years for visual binaries), the inclination $i$ of the orbital plane to be the tangent plane of the celestial sphere at the star, the position angle $\Omega$ (measured from the north through east) of the line of nodes joining the intersection of the orbital and tangent planes, the longitude of periastron $\omega$; the angle between the direction to the ascending node (at which the star crosses the tangent plane while receding from the observer) and that to the point of closest approach of the two stars (periastron). This angle is measured in the orbital plane, in the direction of orbital motion, the major semi-axis of the orbit $a$, usually expressed in kilometers or astronomical units, the eccentricity of the orbit $e$ (a dimensionless number between zero and unity), the time of the periastron passage $T$ (the time which the two stars pass through periastron).

Kowalsky method was described first by [9] and Smart [10] offers formulae and elegant proof. Analytical formulation and computational algorithms for the scheme are presented here briefly. Given the apparent separation $\rho$ in an arc of seconds and the position angle $\theta$ in degree, the equation of the apparent ellipse is given by

$$Ax^2 + 2Hxy + By^2 + 2Gx + 2Fy + 1 = 0, \tag{1}$$

where $x = \rho \cos\theta$ and $y = \rho \sin\theta$.

The constants $A, B, ..., F$ of Equation (1) for the apparent orbit are first derived from the least squares method. After some manipulation one gets the following equations:

$$F^2 - G^2 + A - B = \frac{\cos 2\Omega \, \tan^2 i}{p^2}, \tag{2}$$

where

$$p = a(1-e^2), \tag{3}$$

$$FG - H = -\frac{\sin 2\Omega \, \tan^2 i}{2p^2}, \tag{4}$$

Combining Equations (2) and (4) gives

$$(F^2 - G^2 + A - B)\sin 2\Omega + 2(FG - H)\cos 2\Omega = 0 \tag{5}$$

from which $\Omega$ can be determined.

Using the value of $\Omega$ just found, the value of $\frac{\tan^2 i}{p^2}$ is found by either Equation (2) or Equation (4). Also, we have

$$F^2 + G^2 - (A + B) = \frac{2}{p^2} + \frac{\tan^2 i}{p^2}, \tag{6}$$

But $\frac{\tan^2 i}{p^2}$ has already been determined; hence the value of $p^2$ can be determined from Equation (6). When $p$ has been found, the value of $\tan^2 i$, and hence the inclination, can be calculated. Finally, we can compute $\omega$ from

$$\tan\omega = \frac{(F\cos\Omega - G\sin\Omega)\cos i}{F\sin\Omega + G\cos\Omega} \tag{7}$$

Using the above equations we can determine the geometrical elements ($a$, $e$, $i$, $\Omega$, $\omega$), details of the computational steps will be described in section 2.2. The rest of the seven orbital elements; the period and the time of the periastron passage are computed as follows. The double of the areal constant $C$ computed for the number of the $N$ observed positions ($\rho$, $\theta$) has the form

$$C = \frac{1}{N-1}\sum_{j=1}^{N-1} S_j, \tag{8}$$

where

$$S_j = (x_j \Delta y_j - y_j \Delta x_j)/\Delta t_{j+1}.$$

The period $P$ is then given by

$$P = \frac{2\pi a^2 \sqrt{1-e^2}}{C}\cos i. \tag{9}$$

The mean motion $\mu$ and the eccentric anomaly could be written as

$$\mu = \frac{2\pi}{P}, \tag{10}$$

and

$$E = 2\tan^{-1}\left[\sqrt{\frac{1-e}{1+e}}\tan(v/2)\right], \tag{11}$$

The true anomaly $v$ is computed from

$$\tan(v+\omega) = \tan(\theta-\Omega)\cos i. \tag{12}$$

The mean anomaly $M$ is computed from the Kepler equation

$$M = E - e\sin E. \tag{13}$$

Finally, the time of the periastron passage $T$ is given by

$$T = (\mu t - M)/\mu. \tag{14}$$

### 3. Total and Individual Masses

The total and the individual dynamical masses could be computed using the well-known formula

$$M_a + M_b = \frac{d^3 a^3}{P^2}, \tag{15}$$

where $d$ is the distance to the system in parsecs, $a$ is the semimajor axis in arcsecond and $P$ is the period in years. We determined the mass ratio $M_b/M_a$ from the relation [11]

$$\frac{M_b}{M_a} = 10^{-0.1157(m_b - m_a)}, \qquad (16)$$

$m_a$ and $m_b$ are the apparent magnitudes of the primary and secondary components. Consequently, the individual masses are derived by solving Equations (15) and (16). The errors in the masses could be obtained from the equation

$$\frac{\sigma_M}{M} = \sqrt{9\left(\frac{\sigma_\pi}{\pi}\right) + 9\left(\frac{\sigma_a}{a}\right) + 4\left(\frac{\sigma_P}{P}\right)}. \qquad (17)$$

## 4. Computational Algorithm

To calculate the seven orbital elements, we go through the following computational scheme.

- Input: $\rho$ and $\theta$
1- compute, for $i = 1,...N$.
$$x_i = \rho_i \cos\theta_i.$$
$$y_i = \rho_i \sin\theta_i.$$
2- Solve Equation (1) by the least-squares, yield
$A, B, F, G, H$.

3- Compute the quantities $X_1, X_2, X_3$ from
$$X_1 = -2(FG - H), \quad X_2 = F^2 - G_2 + A - B, \quad X_3 = F^2 + G^2 - (A + B).$$

4- Compute $\Omega$ (the angle of the ascending node) from
$$\Omega = \frac{1}{2} \tan^{-1}[X_1/X_2].$$

5- Compute $Y_1$ and $Y_2$ from
$$Y_1 = X_2/\cos 2\Omega, \quad Y_2 = 2/(X_3 - Y_1)$$

7- Compute the semi-parameter $p$ from
$$p = \sqrt{Y_2}$$

8- Compute $i$ (inclination) from
$$i = \tan^{-1}[p\sqrt{Y_1}]$$

9- Compute $\phi_1$ and $\phi_2$ from

$$\phi_1 = [F\cos\Omega - G\sin\Omega]\cos i, \quad \phi_2 = [F\sin\Omega + G\cos\Omega].$$

10- Compute $\omega$ (the angle of the descending node) from

$$\omega = \tan^{-1}[\phi_1/\phi_2]$$

11- Compute $Y_3$ from

$$Y_3 = \phi_1^2 + \phi_2^2$$

12- Compute $e$ (eccentricity) from

$$e = \sqrt{Y_3}$$

13- Compute $a$ (semi-major axis) from

$$a = p/(1-Y_3^2)$$

14- For all $i = 1(1)N$ compute the true anomalies $v_i$ from

$$v_i = \tan^{-1}\left[\frac{\tan(\theta_j - \Omega)}{\cos i}\right] - \omega; \quad j = 1(1)N.$$

15- For all $i = 1(1)N$ compute the eccentric anomalies $E_i$ from

$$E_i = 2\tan^{-1}\left[\sqrt{\frac{1-e}{1+e}} \tan\frac{v_i}{2}\right]; \quad i = 1(1)N.$$

16- For all $i = 1(1)N$ compute the mean anomalies $M_i$ from

$$M_i = E_i - e\sin E_i, \quad i = 1(1)N.$$

17- Compute the double areal constant from

$$C = \frac{1}{N-1}\sum_{j=1}^{N-1} S_j, \quad S_j = (x_j\Delta y_j - y_j\Delta x_j)/\Delta t_{j+1}.$$

18- Compute the period $P$ from

$$P = \frac{2\pi a^2\sqrt{1-e^2}}{C}\cos i$$

19- Compute the time of periastron passage $T$ from

$$T = (\mu t - M)/\mu,$$

$$t = \frac{1}{N-1}\sum_{i=1}^{N} t_i, \quad M = \frac{1}{N-1}\sum_{i=1}^{N} M_i, \quad T = \frac{1}{N-1}\sum_{i=1}^{N} T_i, \quad \mu = \frac{2\pi}{P}$$

To determine the true anomaly $v$ we go through the following sequence: First, determine the kind of motion, if the motion is retrograde, put, $R1 = v + \omega$. If the motion is direct we put, $R1 = 2\pi - v - \omega$. Then determine the quadrant of $R1$. Second, put $R = \theta - \omega$ and determine the quadrant of $R$. Third, compare between $R$, $R1$ as follow: If the quadrant of $R1$ is greater than the quadrant of $R$ then $v$ must be reduced to $v - \pi/2$ if the quadrant of $R$ is greater than the quadrant of $R1$ then $v$ becomes $v = v + \pi/2$.

## 5. Results

We developed an independent FORTRAN code utilizing the Kowalsky method described in section 2 to calculate the orbits of the four visual binaries whose information is listed in Table 1. The apparent magnitude $m_a$ and $m_b$ are derived from the Washington double star catalog [12], the distance $d$ in parsec is derived from Gaia DR2 [13] except the system WDS J17466-0354, and the spectral types are from [12]. The masses, effective temperatures, and absolute magnitudes are derived from mass-spectral type relation presented by [14]. The position angles and the angular separation are retrieved from the fourth catalog of interferometric measurements of binary stars (astro.gsu.edu/wds/int4.html).

To determine the errors accompanying the orbital elements, we use the trial and error technique, and then the standard deviation is calculated in the usual way. The good orbit is chosen according to the following criteria: The dynamical mass for an orbit may be determined by employing the parallax of Gaia DR2 and compared to the mass derived from the spectral types of the components. Moreover, calculating the ephemerides of from the calculated orbit and determine the orbit with small $\Delta\theta$ and $\Delta\rho$, for this purpose we used the algorithm developed by [15-16].

Table 1: Physical parameters of the four visual binaries.

| WDS/ name | $m_a$ | $m_b$ | $d$ (pc) | Sp. type | Masses | Teff | Mv |
|---|---|---|---|---|---|---|---|
| 02262+3428 | 8.70 | 9.14 | 44.41±0.21 | G8+G9 | 0.99,0.95 | 5559,5450 | 4.92,5.25 |
| 14310-0548 | 8.81 | 8.39 | 41.061±0.43 | G5+G5 | 1.031,1.031 | 5741,5741 | 4.64,4.64 |
| 17466-0354 | 9.34 | 10.22 | ---- | F8 | 1.222,1.222 | 6152,6152 | 3.8,3.8 |
| 12422+2622 | 10.09 | 10.8 | 41.0846 | K4V | 0.8,0.8 | 4400,4400 | 7,7 |

### 3.1 WDS J02262+3428

The visual binary WDS J02262+3428 (HD 15013, HIP 11352, GR2 326940164774368384) is a late type star (G8+G9) with apparent magnitudes $m_a = 8.74$ and $m_b = 9.14$. The Gaia DR2 parallax $\pi = 22.515\, mas$ and the spectral type of the two stars are G5V. The orbit in the six orbit catalog (http://www.astro.gsu.edu/wds/orb6.html) is calculated by [17]. Table 3 lists the two sets of the orbital elements (our orbit and [17] orbit) and the calculated dynamical masses using Equations (15-17). The difference between our time of the periastron passage and that of [17] is remarkable. By investigated the observed angular separation, one can easily get the minimum value of the separation is about 0.057 arcsecond and occurred at the epoch 2000 which ensures our result. The total and the individual mass agree well (within the error) with the masses derived from empirical relation by [14] and listed in Table 1. Besides the observed positions in Figure 1, we illustrate the positions computed from the present orbit and that computed from [17] orbit.

Table 3. Orbit and physical parameters of the visual binary WDS J02262+3428.

| **Element** | **Present work** | **Six orbit catalog** |
|---|---|---|
| $a$ | 0.101±0.003 | 0.099 |
| $e$ | 0.27±0.02 | 0.291 |
| $i$ | 51.47±0.50 | 49.9 |
| $\Omega$ | 13.80±1.5 | 16.0 |
| $\omega$ | 4.13±0.22 | 2.8 |
| $P$ | 6.63±0.30 | 6.937 |
| $T$ | 1998.98±0.20 | 2015.82 |
| $M_t$ | 2.05±0.14 | 1.87 |
| $Ma$ | 1.08±0.07 | 0.99 |
| $Mb$ | 0.96±0.06 | 0.88 |

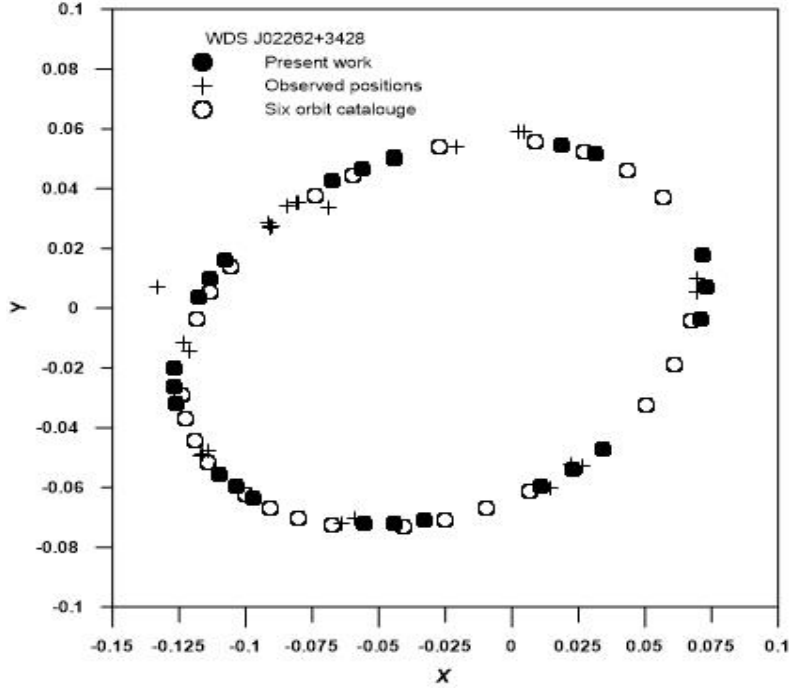

Figure 1: The orbit of the visual binary WDS J02262+3428.

### 3.2 WDS J14310-0548

The two components of the system WDS J14310-0548 (HIP 70973, HD 127352, IDS 14258-0522, Gaia DR2 3641365877340584064) are of late type stars (G5V), the apparent magnitudes are $m_a = 8.81$ and $m_b = 8.39$, the Gaia DR2 parallax $\pi = 24.354$ mas. The orbit in the six orbit catalog is computed by [18]. The total mass computed from the present orbit using the distance derived from the Gaia DR2 catalog (d=41.061 pc) is in good agreement (the mass of the primary is slightly overestimated) with the masses derived from the mass-spectral type relation by [14], $M_a = M_b = 1.03$. In Figure 2 we illustrate the positions computed from the present orbit and that computed from [18] orbit.

Table 4. Orbit and physical parameters of WDS J14310-0548.

| Elements | Present work | Six orbit catalog |
|---|---|---|
| $a$ | $0.24 \pm 0.01$ | 0.243 |
| $e$ | $0.48 \pm 0.02$ | 0.499 |
| $i$ | $50.44 \pm 0.50$ | 49.1 |
| $\Omega$ | $11.62 \pm 0.50$ | 13.8 |
| $\omega$ | $122.82 \pm 0.88$ | 121 |
| $P$ | $21.43 \pm 0.42$ | 22.98 |
| $T$ | $1996.18 \pm 0.60$ | 1993.62 |
| $M_t$ | $2.13 \pm 0.21$ | 1.881 |
| $M_a$ | $1.12 \pm 0.11$ | 0.993602 |
| $M_b$ | $1.01 \pm 0.11$ | 0.884202 |

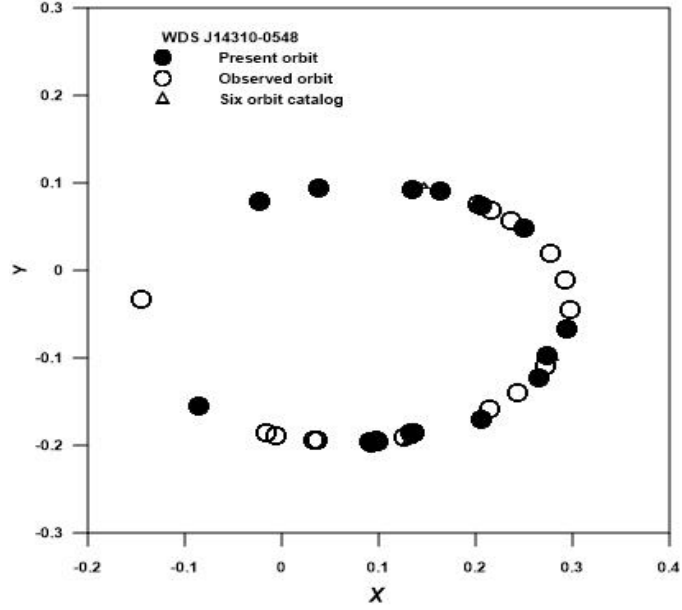

Figure 2: The orbit of the visual binary WDS J14310-0548.

### 3.3 WDS J17466-0354

The spectral type of the system WDS J17466-0354 (ADS 10780, HD 161588, IDS 17413-0352, CCDM J 17466-0354) in Simbad is between F3 and F8 and the apparent visual magnitude is $m_a = 9.34$ and $m_b = 10.22$. There is no parallax for the system in Hipparcos or Gaia DR2 surveys. This neglected system has no orbital elements in the six orbit catalog. We listed the computed orbital elements in Table 5. The computed positions and the observed positions are plotted in Figure 3.

Table 5. Orbit and physical parameters of WDS J17466-0354

| **Element** | **Present work** |
|---|---|
| $A$ | $1.37 \pm 0.03$ |
| $E$ | $0.23 \pm 0.01$ |
| $I$ | $67.58 \pm 0.72$ |
| $\Omega$ | $62.56 \pm 1.71$ |
| $\omega$ | $167.82 \pm 0.40$ |
| $P$ (years) | $56.28 \pm 0.02$ |
| $T$ | $1891.73 \pm 0.05$ |

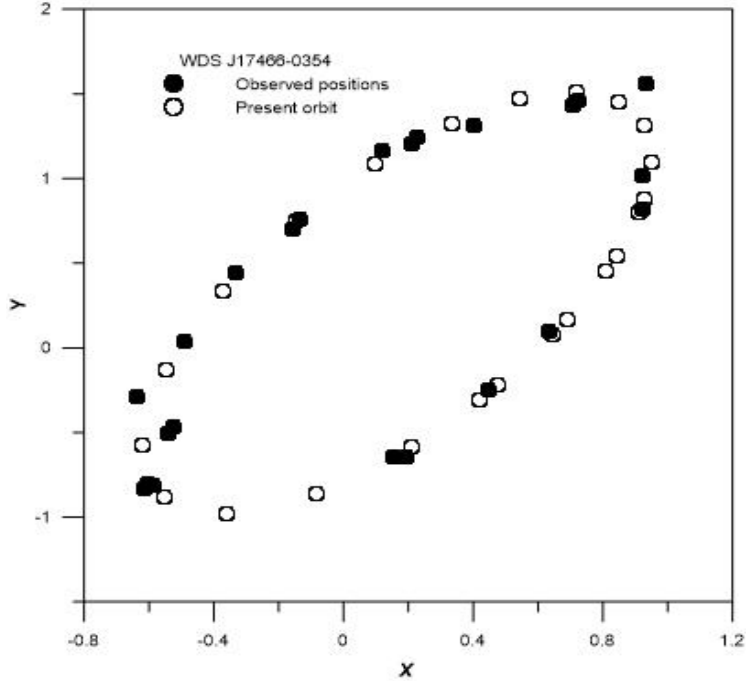

Figure 3: The orbit of the visual system WDS J17466-0354.

### 3.4 WDS J12422+2622

The components of the system WDS J12422+2622 (ADS 8635, HD 110465, HIP 61986) have an apparent magnitudes $m_a = 10.09$ and $m_b = 10.8$. The orbit of the system in the six orbit catalog is calculated by [19]. The total mass computed from the present orbit along with the distance derived from the Gaia DR2 catalog (d=41.0846 pc) is slightly different from that derived from the mass-spectral type relation ($M_a = M_b = 0.8$) [14] but is in good agreement with the massed computed using the orbit of [19].

Table 6. Orbit and physical parameters of WDS J12422+2622

| **Element** | **Present work** | **Six orbit catalog** |
|---|---|---|
| $A$ | $0.41 \pm 0.05$ | 0.415 |
| $e$ | $0.27 \pm 0.01$ | 0.252 |
| $i$ | $31.13 \pm 1.52$ | 26 |
| $\Omega$ | $104.61 \pm 3.42$ | 129.8 |
| $\omega$ | $383.22 \pm 5.38$ | 319.5 |
| $P$ | $63.71 \pm 2.3$ | 61.3 |
| $T$ | $1958.09 \pm 3.21$ | 1959.3 |
| $M_t$ | $1.17 \pm 0.42$ | 1.332 |
| $M_a$ | $0.64 \pm 0.02$ | 0.66 |
| $M_b$ | $0.53 \pm 0.04$ | 0.66 |

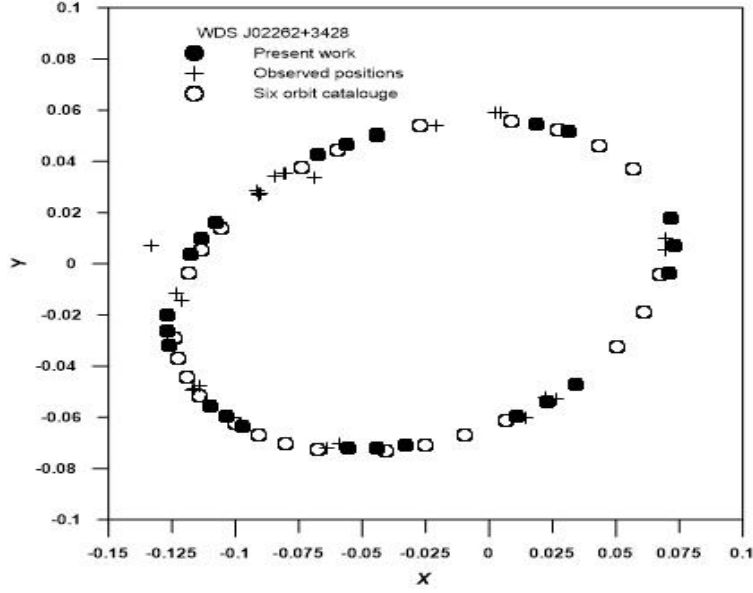

Figure 4: The orbit of the visual system WDS J12422+2622.

## 6. Conclusion

In the present work, we introduced a computational algorithm to determine the orbital elements of the visual binaries. The geometrical elements are computed using the Kowalky method, while the period and the time of the periastron passage are determined by utilizing the double areal constant. We derived the orbital elements, total masses, and individual masses for the four visual binaries WDS J02262+3428, WDS J14310-0548, WDS J17466-0354, and WDS J12422+2622. The comparison between our orbits with that listed in the six orbit catalog gives a good agreement. Comparison between the total dynamical masses derived from the present orbits and that from the empirical mass-spectral type relation gives a good agreement. We introduced a new orbit for the neglected visual binary WDS J17466-0354.

**Acknowledgments**

The author gratefully acknowledges the approval and the support of this research study by the grant number Sci/2019/1/10/F/8282 from the Deanship of Scientific Research at Northern Border University, Arar, Saudi Arabia.